# Hot spot-mediated non-dissipative and ultrafast plasmon passage

Eva-Maria Roller[1*], Lucas V. Besteiro[2*], Claudia Pupp[1], Larousse Khosravi Khorashad[2], Alexander O. Govorov[2†], Tim Liedl[1†]

*[1]Faculty of Physics and Center for NanoScience, Ludwig-Maximilians-Universität München, 80539 Munich, Germany*
*[2]Department of Physics and Astronomy, Ohio University, Athens, Ohio 45701, USA*

[†] govorov@helios.phy.ohiou.edu; tim.liedl@lmu.de;

**Plasmonic nanoparticles hold great promise as photon handling elements and as channels for coherent transfer of energy and information in future all-optical computing devices.[1-5] Coherent energy oscillations between two spatially separated plasmonic entities *via* a virtual middle state exemplify electron-based population transfer, but their realization requires precise nanoscale positioning of heterogeneous particles.[6-10] Here, we show the assembly and optical analysis of a triple particle system consisting of two gold nanoparticles with an inter-spaced silver island. We observe strong plasmonic coupling between the spatially separated gold particles mediated by the connecting silver particle with almost no dissipation of energy. As the excitation energy of the silver island exceeds that of the gold particles, only quasi-occupation of the silver transfer channel is possible. We describe this effect both with exact classical electrodynamic modeling and qualitative quantum-mechanical calculations. We identify the formation of strong hot spots between all particles as the main mechanism for the loss-less coupling and thus coherent ultra-fast energy transfer between the remote partners. Our findings could prove useful for quantum gate operations, but also for classical charge and information transfer processes.**

Low-dissipative transfer of excitations over short and long distances is at the heart of information science as well as energy harvesting. Energy transfer processes, for example, play a key role in highly efficient dipolar interactions in the light harvesting complexes of chloroplasts and low-loss exciton transport is sought after in solar cell development. Both Förster-Resonance Energy transfer (FRET) in biomolecular systems and diffusion of excitons in solar cells are incoherent processes and therefore often dissipative. Also information inside computer chips is, as of today, processed and transferred incoherently. On the other hand, coherence is a crucial feature in interferometry and indispensable in future quantum computation.

With the goal to achieve fast and coherent transfer between nanoscale components, a variety of quantum approaches emerged. Generally, quantum mechanical tunneling is fast enough to avoid inelastic scattering during the passage, which makes it a low-loss process that is widely used in modern electronic technology. In particular, chains of potential wells were proposed for tunneling by adiabatic passage (CTAP) [1,2] and optical stimulated Raman adiabatic passage (STIRAP) has been realized by transferring spin populations of two long living quantum states *via* an optical third state.[11,12]

Another approach for transport of information at the nanoscale is the use of plasmonic components.[13] Plasmons are coherent, but they are usually dissipative and have only short life times. Both disadvantages are the result of the high scattering probability of electrons in metals, which hampers the use of plasmonic waveguides for transfer applications. Inspired by efforts to exploit quantum mechanical mechanisms we now propose a particle trimer system where plasmons are coherently transferred over extended distances. The system consists of two identical but spatially separated nanoparticles of one type and a third intermediate nanoparticle of a different type, the latter exhibiting also a different energetic level. The plasmons are transferred between the identical nanoparticles *via* the third nanoparticle, even with the two identical particles being too far away from each other to couple directly.

The experimental realization of such a plasmon-based transfer system requires full control over spatial organization of heterogeneous nanoparticles. DNA-based self-assembly offers the possibility to fabricate nanoscale objects that can accommodate inorganic particles at extremely well defined positions with high yields.[6-9,14-17] Previous assemblies with DNA-based templates consist of arrays,[8] chain-like,[18,19] helical,[9] or ring shaped[20] arrangements of metal nanoparticles, as well as chains of dyes[21] and quantum dots.[22,23] Owing to the possibility of functionalizing particle species with orthogonal DNA sequences – i.e. sequences that do not interfere with each other – heterogeneous particle architectures have been realized including dimers and trimers,[23,24] lattices,[8] core-satellites[15,25] and rings[20]. Lithographical attempts to build plasmonic devices from heterogenous metals consist of bimetallic nanodot arrays[26] and nanoantenna dimers of gold and silver disks[27]. However, top-down approaches suffer from limited spatial control on the scale below tens of nanometers and generally result in less homogenous crystalline structures made of sputtered or epitaxially deposited materials. Colloidal nanoparticles, in contrast, exhibit high crystalline quality and sharp size-distributions.

To overcome limitations of top-down lithography, we here use a DNA origami structure to spatially arrange gold and silver nanoparticles (AuNPs and AgNPs) in hetero-trimers with nanometer-precision and high assembly yields.[28,29,30] Our DNA origami template consists of a cylindrical 14-helix bundle (for design details and assembling procedures see Methods section and Supplementary Information fig. S1), which offers lengthwise three equally spaced sequence-specific attachment sites for DNA-functionalized nanoparticles. AuNPs functionalized with a DNA sequence complementary to the outer sites and one AgNP functionalized with a sequence complementary to the middle site were hybridized to each origami template (Fig.1a). The resulting heterogeneous nanoparticle trimer displays a designed interparticle gap of 40 nm between the two outer AuNPs with the AgNP accommodated in between. After assembly, we confirmed this configuration by transmission electron microscopy (TEM) (Fig. 1b and 1c, Methods section and Supplementary Information fig. S2). Note that the AgNPs in our experiments are slightly smaller than the AuNPs and exhibit less contrast.

Conceptually, the distance between the AuNPs alone is too large to support plasmonic coupling and thus no transfer of energy is expected. The AgNP bridges this gap and transfers energy coherently between the two outer AuNPs and serves as a loss-less virtual transmitting state (Fig 1d). This can be understood as follows: If we excite our system in the gold plasmon resonance, the two gold plasmons are in resonance, however, at a distance too long to couple. At the same frequency, the silver plasmon is not in resonance but becomes involved as a quasi-resonant virtual state that operates as a transmitter. Since silver has a relatively small Drude dissipation constant, Ag plasmons have a narrow plasmon peak and exhibit a very strong induced dipole moment even for relatively small NP sizes. These important features make AgNPs excellent transmitter elements that allow us to connect the two gold plasmons almost without dissipation as we show in this study using both experiment and theory.

We characterized the plasmonic coherent transfer system with dark-field scattering spectroscopy of individual hetero-trimers immobilized on a glass substrate in air and compared the resulting spectra with theoretical calculations (see also Methods section and Supplementary Information). Figure 2a displays the scattering spectra of several AuNP-AgNP-AuNP trimers in comparison to an Au-Au homo dimer, missing the middle AgNP. The dominant resonance wavelength shifts from 549 nm for the AuNP homo dimer to 586 nm for the hetero-trimer structure. At the same time the peak intensity increases by a factor of ~ 4. Apparently, the middle AgNP indeed serves as a connector to enable strong coupling between the two outer AuNPs. The experiments further show that the use of larger AgNPs results in a stronger dominant peak intensity at 586 nm. Concurrently, the resonance level of the AgNP becomes visible as a small peak at 445 nm.

The particle geometries in our numerical simulations were adjusted to the particle positions and sizes as determined from SEM and TEM images. For the simulation of the dimer structure, two 40 nm AuNPs separated by a 38 nm gap were chosen. For the trimer simulations a 30 nm AgNP was placed in the gap (see also Supplementary Information figs. S3, S4). Figure 2b shows that the simulated scattering cross sections are in excellent agreement with the observed spectra. For the hetero-trimer the dominant plasmon resonance peak shifts ~ 40 nm to the red. Furthermore, the peak intensity is increased by a factor of four compared to the homo-dimer. Most importantly, the simulations confirm that the AgNP does not dissipate energy at the resonance mode of the hetero-trimer, but instead transfers the energy coherently between the two outer AuNPs (Fig. 2c).

Polarization-resolved scattering measurements give additional insight into this non-dissipative passage. Figure 3a shows such measurements for a single hetero-trimer. For the linear detecting polarizer set to 90° with respect to the long trimer axis only small transversal resonance peaks of the single AgNP and AuNPs are visible at ~ 450 nm and ~ 550 nm, respectively. At parallel (0°) polarization, the coupled, red-shifted mode appears. Our simulations fully reproduce these observations (Fig. 3b).

The charge color maps in Figure 3c display the key feature of the plasmon coupling in our structures: The energy transfer between the Au-NPs occurs *via* the plasmonic hot spots formed in the Au-Ag gaps (see also Supplementary Information fig. S5). These highly localized spots generate surface charges involving high multipole harmonics, which contrasts the standard picture for energy transfer processes between, e.g., dye molecules.

By analyzing the data in Figures 2a and b, we can estimate the lifetime of the plasmon in the trimer and its transfer time between the two Au NPs. From the width of the plasmon peak we obtain a lifetime of ~ 8 fs. From the splitting of the L-plasmon resonances ( $\Delta_{splitt}$ ) we calculate that $\tau_{transfer} \sim \pi / (2 \cdot \Delta_{splitt}) \sim 4.7\,\mathrm{fs}$ (see Supplementary Information notes and fig. S6). Exact electromagnetic simulations of the plasmon transfer dynamics support this analytical estimate. For this we placed an exciting dipole sending an 8.3 fs long Gaussian pulse near one of the AuNPs and observed the dipolar moments of both AuNPs over time (Fig. 4). The electric dipole moments and the energies stored on the AuNPs show time-oscillating traces and thus coherent transfer *via* the AgNP with a characteristic transfer time of ~ 5 fs (Fig. 4 and Supplementary Information figs. S7, S8). No such oscillations occur in the absence of the middle particle (Supplementary Information fig. S9). As expected, the energy dissipation in the central transmitting AgNP is small which shows the low-loss mechanism of information and energy transfer between the AuNPs mediated by the virtual Ag-plasmon. In summary, the observed plasmon lifetime and transfer time are both on the femtosecond time scale enabling efficient and coherent transfer. Due to the strong coupling of the plasmonic dipoles, this process is orders of magnitude faster than FRET, which only achieves ps-times both in plants and between FRET dyes in the lab.[31] In principle, FRET in molecules and semiconductor nanocrystals can be accelerated by plasmons, but only at the cost of high losses in the metal component.[32]

As our structure is describable as a system of coupled oscillators, we can also apply a quantum model of the plasmonic excitations that should yield similar qualitative results. In this model the plasmons in the three NPs are considered as three quantum oscillators that are coupled by Coulomb forces and have each three degrees of freedom (Fig. 5a).[33] To keep the calculations simple, we assume the dipolar limit for the plasmons, which, of course, underestimates the strength of coupling in our real samples. The total number of modes is 9 and the modes can be bright (B) or dark (D) and longitudinal (L) or transverse (T). They further exhibit certain degeneracies (see Fig. 5a and Supplementary Information for details and model parameters). The Hamiltonian of the coupled plasmonic oscillators reads

$$\hat{H} = \sum_{\alpha} \hbar\omega_{p,\alpha} \left( \hat{c}_{\alpha}^{\dagger} \hat{c}_{\alpha} + \frac{1}{2} \right) + \hat{H}_{\text{int}}, \qquad (1)$$

where $\alpha(\beta)$ are the indices of all possible plasmonic states of isolated NPs. The quantum index can be represented as $\alpha = (i, \gamma)$, where $i = 1,2,3$ is the NP number and $\gamma = x, y, z$ is the direction of oscillation. In Eq. 1, $\omega_{p,\alpha}$ and $\hat{H}_{\text{int}}$ are the plasmon frequencies of isolated NPs and the Coulomb coupling operator, respectively. The Hamiltonian (1) can be easily diagonalized and the spectrum of vibrations can be found. The most interesting modes are the collective longitudinal ones:

$$\omega_{LB,Au-like} = \omega_{p,Au} - \Delta_{LB,Au}, \; \omega_{LD,Au-like} = \omega_{p,Au} + \Delta_{LD,Au}, \; \omega_{LB,Ag-like} = \omega_{p,Ag} + \Delta_{LB,Ag} \quad (2)$$

where the positive parameters $\Delta$ are the energy shifts that appear in the spectrum due to the Au-Au and Au-Ag plasmonic interactions.

This relatively simple quantum model reproduces qualitatively the main important features of our experimentally and computationally obtained spectra (Fig. 5b), which makes this one of the rare examples where classical and quantum pictures are equivalent when considering the coherent properties of coupled plasmons. In particular, the quantum model reproduces the red shift of the main Au-like L-plasmon, the appearance of splittings in the spectra, and a larger splitting for the L-modes as compared to the T-modes. Again, we observe that the Ag-plasmon plays the role of a mediator for the enhanced coupling between the two Au-plasmons. Despite the drastic simplification of our quantum model, which ignores multipolar modes, the formation of hot spots and some other fine details of the system, it helps to understand all plasmonic modes and the characteristic spectral shifts.

Our heterogeneous particle chain where, in contrast to conventional homogeneous mono-metallic waveguides, a silver nanoparticle is introduced as a coherent transmitter allows for ultrafast excitation transfer with almost no losses in the transmitting element. Advanced experimental methods such as nonlinear time-resolved fs-spectroscopy in a pump-probe setup could in future studies record the coherent plasmonic dynamics of the Au-Ag-Au trimer. In a nonlinear plasmonic regime and employing methods of quantum spectroscopy, such a trimer can become a model system to create and control pairs of plasmonic quanta (qubits) and their quantum entangled states as it is currently done for photons and electrons.

**Acknowledgements:** This work was funded by the Volkswagen Foundation, the DFG through the Nanosystems Initiative Munich (NIM), through the ERC Starting Grant ORCA (GA N°:336440). A.O.G. and L.V.B acknowledge additional support from the US Army Research Office (W911NF-12-1-0407).

**Author Contributions:** *These authors contributed equally to this work. E.M.R., A.O.G. and T.L. conceived the experiments and co-wrote the manuscript. E.M.R. designed the structure and analyzed the data. E.M.R. and C.P. performed the experiments. L.V.B. and L.K.K. performed the simulations, A.O.G developed the quantum model. All authors contributed to the interpretation and general discussion and reviewed the manuscript.

**Author Information**: The authors declare no competing financial interests. Correspondence and requests for materials should be addressed to TL (tim.liedl@lmu.de).

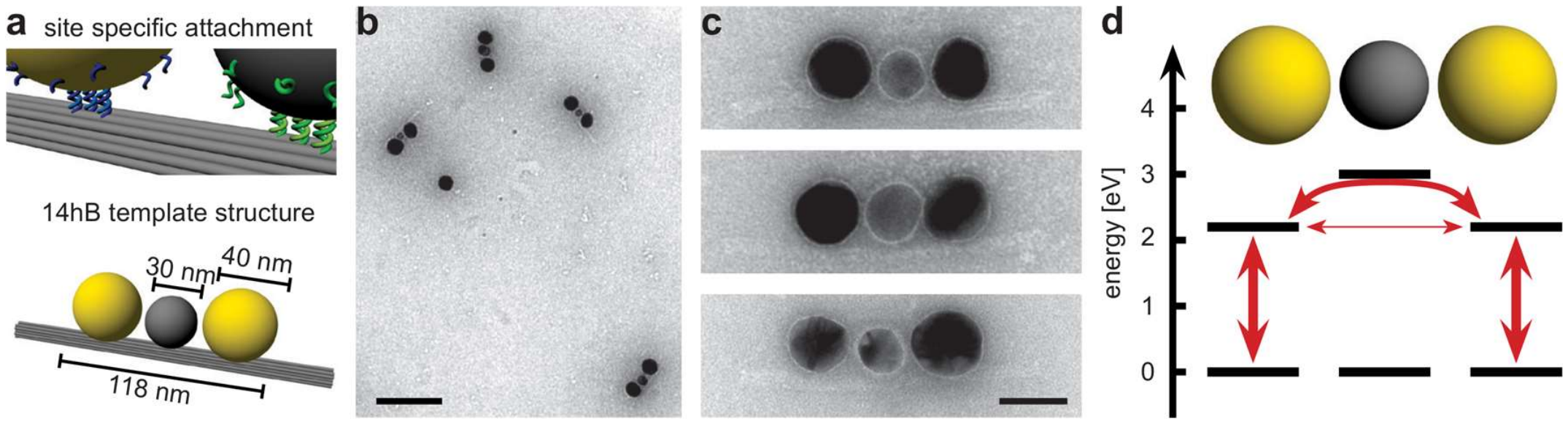

**Figure 1 | Plasmonic transfer system. a,** Design scheme of the DNA origami-templated heterogeneous trimer structure. AuNPs and AgNPs bind *via* specific DNA sequences (shown in blue and green) to designated sites on the DNA origami structure. **b,** Wide field TEM image of the assembled hetero-trimers. Scale bar: 200 nm. **c,** High magnification TEM images reveal the two outer AuNPs, the middle AgNP and the DNA origami scaffold. Scale bar: 40 nm. **d,** Energy scheme and concept of the plasmonic transfer system. The middle Ag particle serves as a virtual transmitter connecting the two separated Au particles.

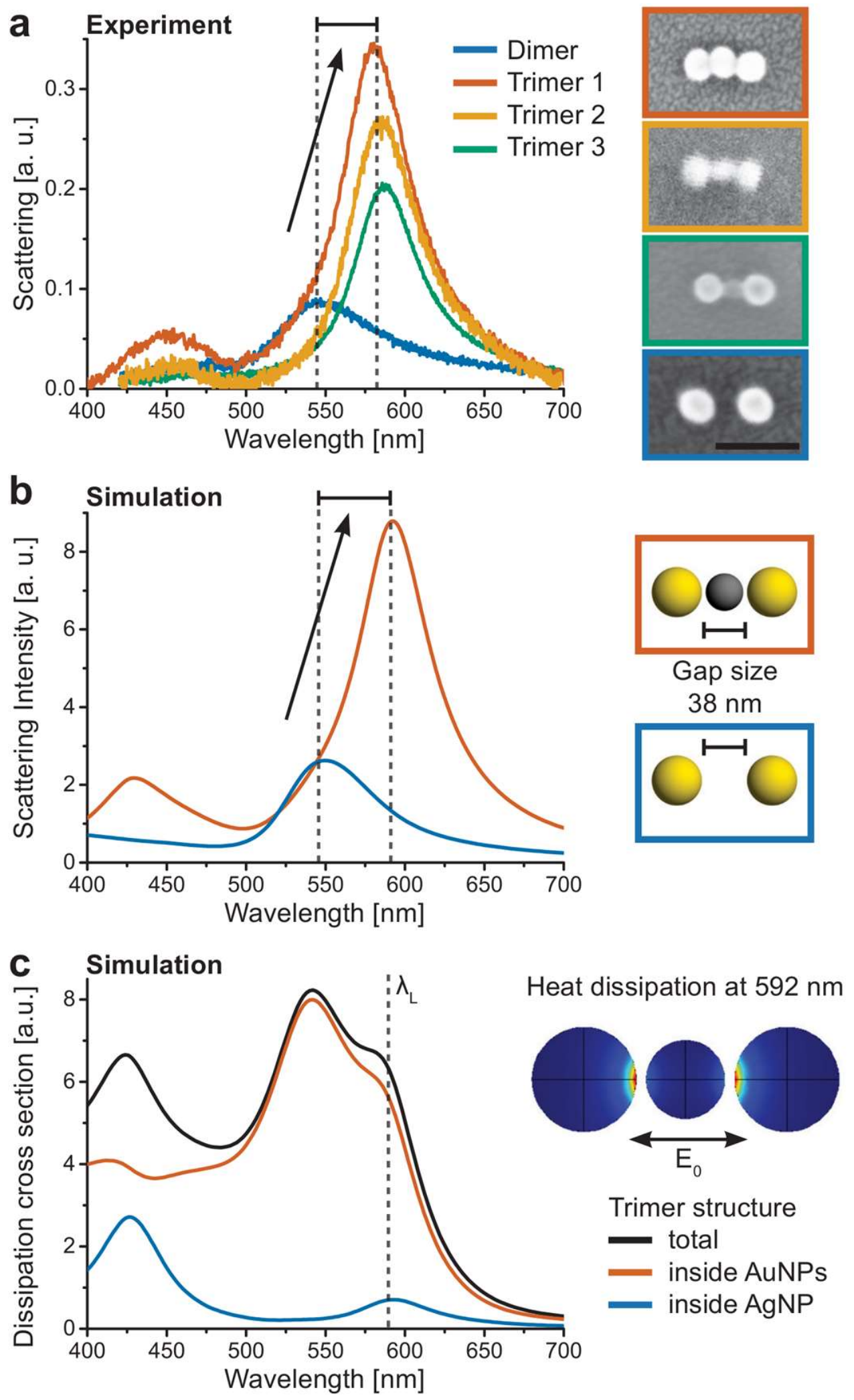

**Figure 2 | Dark-field scattering spectra and simulations of the plasmonic transfer system. a,** Single structure scattering spectra of one homo dimer and three hetero-trimers and SEM images of the corresponding structures. Scale bar: 100 nm. A plasmon resonance shift of ~ 40 nm is observed. **b,** Numerical simulations for a homo dimer structure of two 40 nm AuNPs with a 38 nm gap (blue) and for the same configuration with a 30 nm AgNP placed in the gap (orange). **c,** The simulated heat dissipation cross section of the hetero-trimer reveals that the AuNPs are dissipative at the resonance mode at $\lambda_L$ = 592 nm, while the AgNP is not. The color map shows the local heat dissipation in the longitudinal mode. Most of the dissipation appears in the hot spots on the AuNPs.

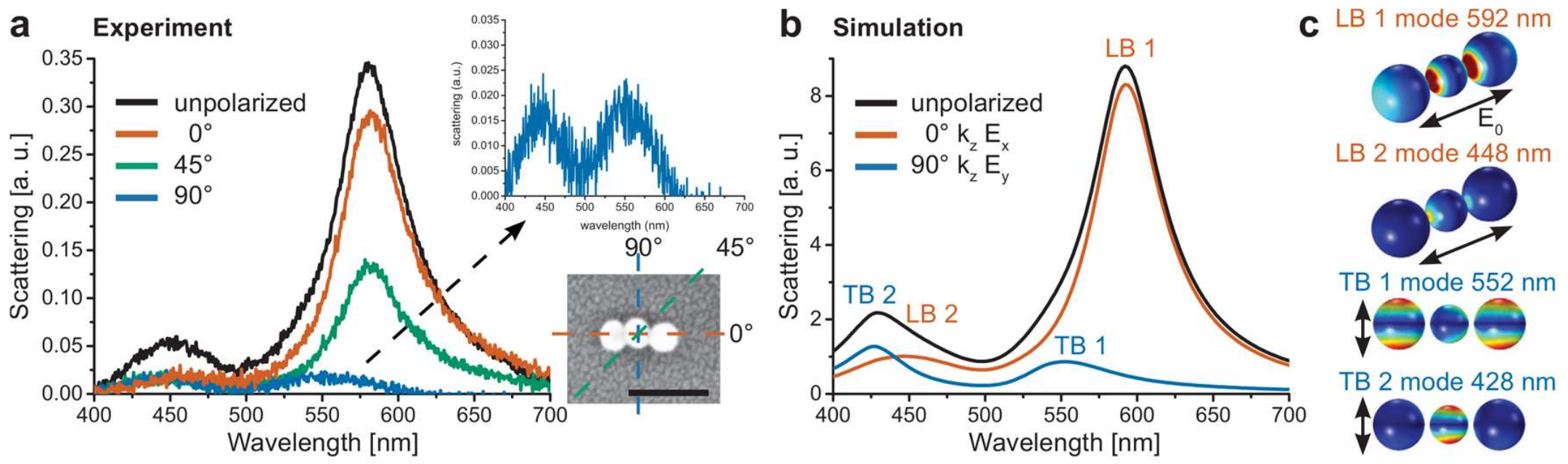

**Figure 3 | Polarization-resolved measurements and calculations. a,** Polarization-resolved scattering spectra of a hetero-trimer structure. Only the resonances of uncoupled AuNPs and single AgNPs are detectable at 90° orientation of the polarizer in the detection path. The resonant mode appears in a parallel orientation of the polarizer. Inset: Orientations of the polarizer in respect to the trimer. Scale bar: 100 nm. **b,** Simulated polarization-resolved scattering spectra and **c,** surface charge maps corresponding to the longitudinal and transversal bright modes (LB, TB) are in excellent agreement with the experimental results.

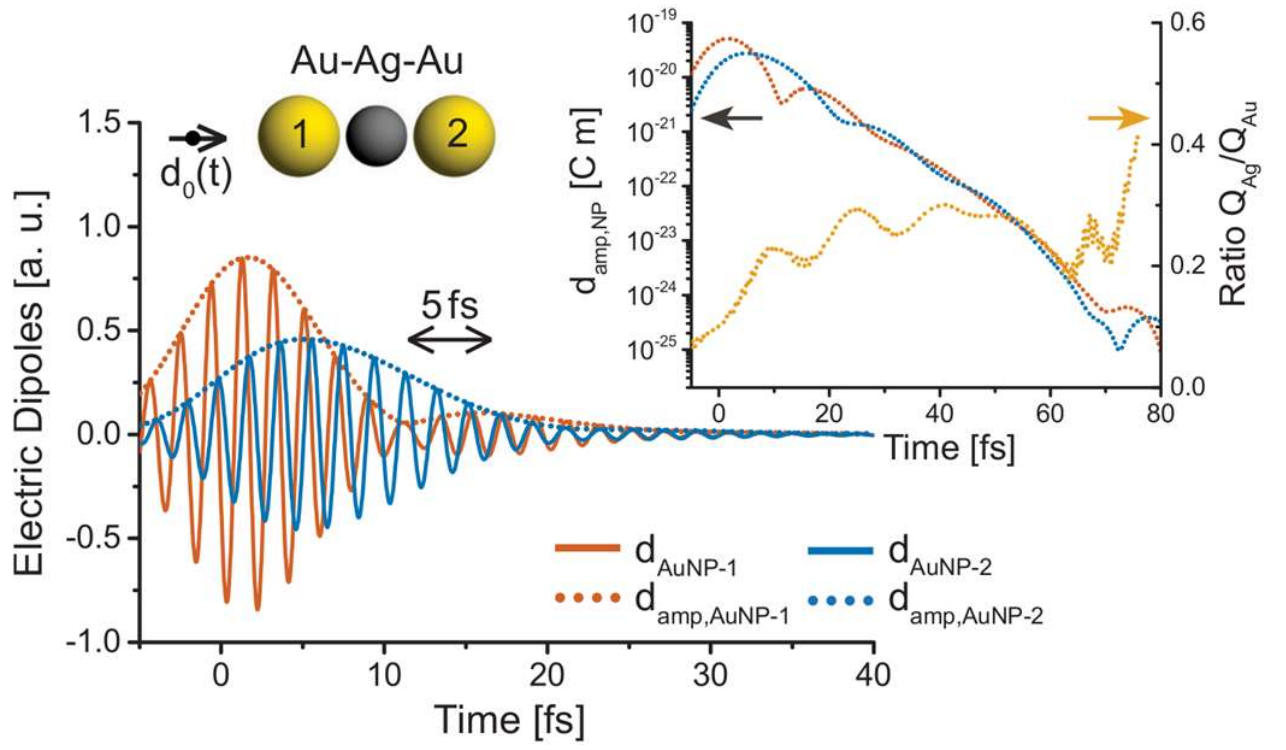

**Figure 4 | Full electromagnetic simulations of the plasmon transfer dynamics. a,** We placed a dipole sending a short Gaussian excitation pulse with a duration of 8.3 fs (FWHM) and a central frequency of $\hbar\omega = 2.21 eV$ closely to one AuNP and monitored the dipoles in both AuNPs. Solid lines show the dipole oscillation in the left (orange) and in the right (blue) particle, dotted lines indicate the time variation of the amplitudes (envelope functions) of the corresponding NP dipoles. Inset: Dipole oscillations in both particles (blue and orange lines) and the ratio $Q_{Ag}$ / $Q_{AU}$ between the dissipations in the two AuNPs and in the mediating AgNP (yellow line). The dissipation in the Ag mediator is small at all times. Oscillations between both Au particles indicate coherent transfer with an estimated transfer time of 5 fs.

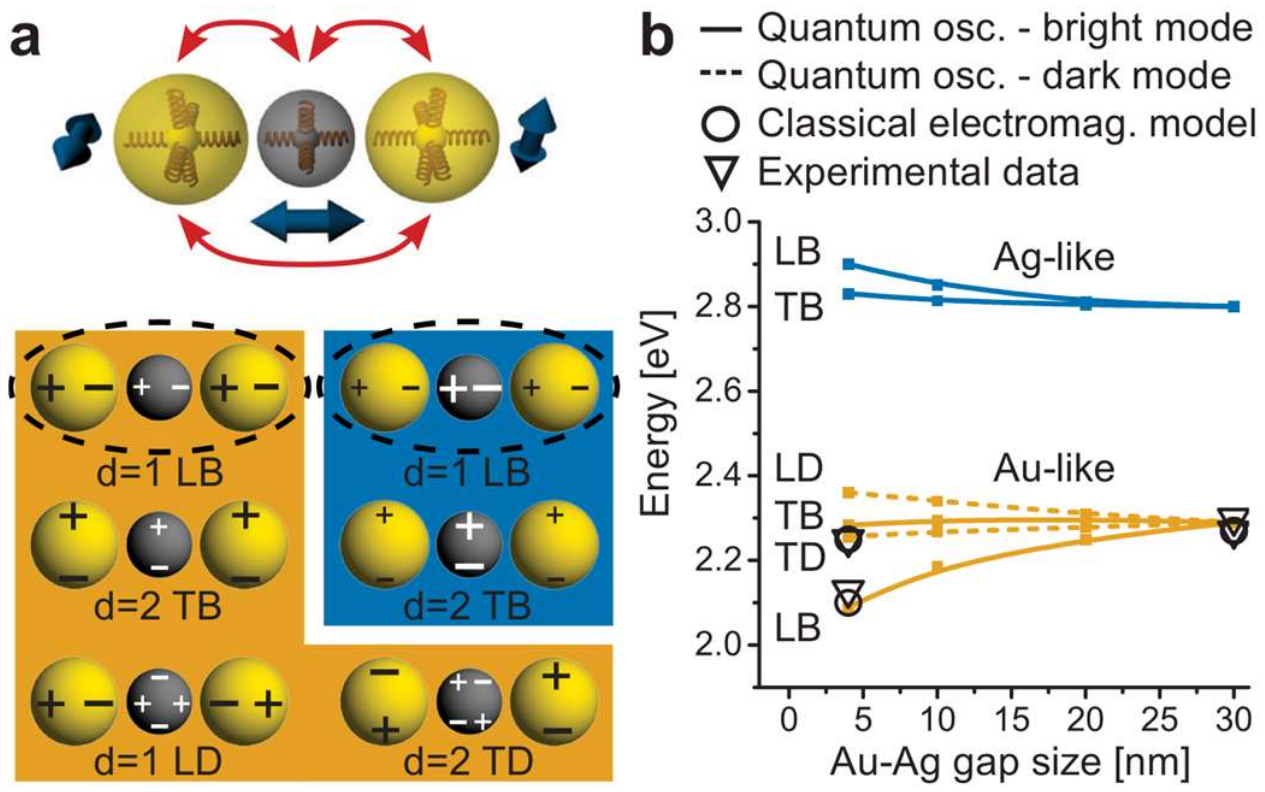

**Figure 5 | Quantum model. a,** Scheme displaying the harmonic-oscillator model of the plasmonic passage (top) and classification of the plasmonic modes (bottom). The main contributing mode to the plasmon passage and the mode assisting the passage of Au plasmons are circled with dashed lines. The label d denotes the degeneracy of the quantum plasmonic states. **b,** Energies of the different plasmonic modes calculated *via* the quantum oscillator model (solid and dashed lines), simulated by the classical electromagnetic model (circles) and experimentally obtained bright modes (triangles).

## Methods

### Assembly of heterogeneous particle trimer structures.

*DNA origami folding.* In DNA origami, a ~ 8000 nucleotide long viral single-stranded DNA (ssDNA) scaffold is folded into a programmed shape with the help of ~ 200 short, synthetic ssDNA staple strands. Due to the sequence-defined assembly of the DNA structure the location of each staple strand and each DNA base within the structure is exactly defined. By extending a selected subset of the staple strands with tailored anchor sequences, the folded DNA origami structure can be employed as a breadboard exhibiting unique and sequence-specific binding sites for DNA-modified metal nanoparticles. Our DNA origami structure offers two outer attachment sites with the same DNA anchor sequence and a middle site with an orthogonal sequence. The DNA origami 14 helix bundle (14hB) was folded using 10 nM of the scaffold p8634, 100 nM of each staple strand, 10 mM Tris, 1 mM EDTA (pH 8) and 16 mM $MgCl_2$. This mixture was heated to 65 °C for 20 min and then slowly cooled down to 20 °C over a period of 40 hours. Specific staple strands at the attachment site were elongated by either 15 x A bases for AuNP attachment or by the sequence ATG TAG GTG GTA GAG AA for AgNP attachment. Thus each attachment site consists of five single-stranded extensions of staples whose ends are located close to each other on the surface of the origami structure. The extended staple strands are labeled in Figure S1 with red color. All the staples for nanoparticle attachment *via* DNA hybridization were already included into the folding solution according to the targeted final NP-DNA origami structure (trimer or dimer structure). After folding the structures, a purification step with a 1 % agarose gel in 1 x TAE (40 mM Tris, 40 mM Acetic Acid, 1 mM EDTA, pH 8) containing 11 mM $MgCl_2$ was performed. The band containing the structures was cut out from the gel with a razor blade. The DNA origami structures were recovered with a pipette while squeezing the gel band between two glass slides. The concentration of the 14-helix bundle after purification was determined via UV-Vis spectroscopy (Nanodrop).

*Concentration of AuNPs and conjugation with DNA.* First, 40 nm AuNPs (BBI Solutions, 20 ml) were concentrated using the protocol of Schreiber *et al.*[12] The AuNPs were mixed with 8 mg BSPP (Bis(*p*-sulfonatophenyl)phenylphosphine dihydrate dipotassium salt, Sigma-Aldrich) and shaken for 3 days. Afterwards NaCl was added till a color change to blue was observed. Then the solution was centrifuged at 1,600 rcf for 30 min and the supernatant discarded. Next, 1 ml of 2.5 mM BSPP in H2O and 1 ml Methanol was added. After vortexing, the solution was centrifuged again at 1,600 rcf for 30 min and the supernatant was discarded. The concentrated AuNPs were redissolved in 200 µl 2.5 mM BSPP and their concentration was determined *via* UV-Vis spectroscopy (Nanodrop). The following functionalization of the AuNPs with 5' thiol-modified ssDNA strands (Biomers.net, 19 x T bases) has two goals: First, the DNA coverage of the AuNPs renders them stable against high $MgCl_2$ concentrations as they are used within the DNA origami folding process. Second, the sequence of the thiol modified ssDNA strands is chosen to be complementary to the single-stranded extensions of the staple strands that together form the AuNP attachment sites on the 14-helix bundle structure. Thus the AuNPs can hybridize to the predesigned sites on the DNA origami structure. For the functionalization a ratio of AuNPs:thiol ssDNA of 1:5,000 was used. 0.5 x TBE buffer was added to the AuNPs and the ssDNA and the solution was kept on a shaker for 3 days. Afterwards a purification step was performed to get rid of the unbound ssDNA strands. For that the AuNPs were run over a 100 kDa MWCO centrifugal filter (Amicon Ultra, Millipore, 5 min, 8,000 rcf) followed by additional 8 centrifugation steps with a filter exchange after 4 steps. This purification from unbound ssDNA strands is crucial to avoid blocking of the attachment sites on the DNA origami structure by free complementary ssDNA strands. Best yields of AuNPs-to-DNA-origami-binding can be achieved if the last centrifugation steps are performed directly before mixing the AuNPs with the DNA origami structures.

*AgNPs functionalization with DNA.* For the functionalization of AgNPs with ssDNA a 5' thiol-modified sequence TTC TCT ACC ACC TAC AT (biomers.net) was used. A functionalization sequence that is different to the AuNP sequence guarantees that the attachment of AuNPs and AgNPs to the 14-helix bundle is specific. First, the as-purchased AgNPs (Cytodiagnostics, 50 nm, 1 ml) were mixed with the sequence (40 µl, 100 µM stock concentration) and 1 x TE buffer. This solution was kept protected from light on a shaker for one week. Afterwards NaCl aging was performed by slowly raising the NaCl concentration to 100 mM over the course of one day. Then short 5'thiol-modified ssDNA (5x T, MWG eurofines, 20 µl, 1 mM stock concentration) were added. This short DNA strand is used as "back filler" to assure high $MgCl_2$ stability of the AgNPs while it is too short to hybridize to the DNA origami structure. The solution was left again over night for incubation. Then the NaCl concentration was raised to a final concentration of 500 mM over the course of 6 hours. As the last step, the DNA-functionalized AgNPs were purified from excess unbound ssDNA strands by using 100 kDa MWCO centrifugal filters as described in the AuNPs procedure.

*Functionalization of DNA origami 14-helix bundle structure with AuNPs and AgNPs.* After determining the concentration of the DNA-modified AgNPs and AuNPs *via* UV-Vis spectroscopy, they were mixed together with the purified DNA origami 14-helix bundle template. This structure offers two attachment sites for AuNPs and one attachment site for AgNPs. Each attachment site consists of five elongated staple strands offering the complementary sequence to the corresponding metal nanoparticle. The DNA origami structure and the metal nanoparticles were mixed in a ratio attachment site:AuNP/AgNP of 1:4. The excess of metal nanoparticles over attachment sites ensures high binding yields and prevents crosslinking of origami structures mediated by nanoparticles. After overnight incubation of the solution, a 0.7% agarose gel electrophoresis in 1 x TAE and 11 mM $MgCl_2$ buffer was run to separate the desired heterogeneous trimer structures from excess, unbound metal nanoparticles and from aggregates (Figure S2). The band containing the trimer structures was excised from the gel and the nanostructures extracted by squeezing the cut gel band between two glass slides. The solution received in a pipette contained the purified final trimer structures.

**Characterization of heterogenous particle trimer structures.**

*Transmission electron microscopy (TEM).* TEM was used to control the correct assembly of the particle trimer nanostructures. A droplet of the solution containing the purified structures was deposited on a plasma-exposed carbon-formvar-coated TEM grid (Ted Pella) and then dabbed off after 3 minutes. The grid was stained with 1% uranyl formate for 15 seconds. Imaging was performed with a JEOL JEM-1100 at an acceleration voltage of 80 kV.

*Dark-field scattering spectroscopy.* In order to take scattering spectra of single trimer structures the purified trimer solution was diluted 20 x in 1 x TAE buffer containing 11 mM $MgCl_2$ and immobilized on cleaned glass cover slides. Therefor, a droplet of the diluted solution was deposited for 5 min onto the glass slides, and then ddH2O was rinsed over the slide to wash away salt residues. To avoid denaturation of the DNA structures, the slides were dried with a nitrogen flush immediately. The measurements were then performed in air. The dark-field scattering spectra were collected with a home-build dark-field set up in transmission mode using a 100x air objective (Olympus) and an oil condenser (Olympus NA 1.4) with a 100 W halogen bulb as illumination source coupled to an Acton SP2300 spectrometer (Princeton Instruments). Polarization-resolved scattering spectra were taken by exciting the system with unpolarized incident light and detecting the scattered light through a rotatable polarizer. Scattering spectra of single AuNP dimer structures (no intermediate AgNP) and single AuNP-AgNP dimer structures[34] (only one outer AuNP) were performed accordingly. The spectra are shown in the Supplementary Information fig. S3 and fig. S4.

*Scanning electron microscopy (SEM).* The single structures characterized in the dark-field set-up were further analyzed by SEM. Therefor, the glass slides were subsequently sputtered with a 3 nm gold palladium layer and imaged using a Gemini Ultra Plus field emission SEM (Zeiss). The images were taken using the in-lens detector and an electron acceleration voltage of 2 kV at a working distance of 3.0 mm.

**Data availability.** The data that support the plots within this paper and other findings of this study are available from the corresponding author upon reasonable request.

# Supplementary Information

# Hot spot-mediated non-dissipative and ultrafast plasmon passage

Eva-Maria Roller[1*], Lucas V. Besteiro[2*], Claudia Pupp[1], Larousse Khosravi Khorashad[2], Alexander O. Govorov[2†], Tim Liedl[1†]

[1]*Faculty of Physics and Center for NanoScience, Ludwig-Maximilians-Universität München, 80539 Munich, Germany*
[2]*Department of Physics and Astronomy, Ohio University, Athens, Ohio 45701, USA*

[†] govorov@helios.phy.ohiou.edu; tim.liedl@lmu.de;

**Supplementary Notes.**

**Theoretical Modelling of trimer plasmonic bus structures.**

Classical theoretical simulations of the NP complexes. Theoretical computations of the NP chains were performed using classical electrodynamics with Comsol Multiphysics with standard boundary conditions and with local dielectric functions for the gold sphere[2] and the silver sphere.[3] The optical dielectric constant of the matrix was taken as an average between water and glass ($\varepsilon = (1.8 + 2.5) / 2 = 2.15$). The trimer particle chain was simulated with the following parameters: diameter of Au sphere 40 nm, diameter of the Ag sphere 30 nm and gap size in between the spheres 4 nm. The scattering intensities were simulated with light approaching the plane of the trimer at an angle of 20° and the elastic scattering was collected computationally on the vertical axis. Calculations were performed for incident beams with two orthogonal polarizations and the resulting scattering spectra were averaged over the incident polarizations.

The local dissipation spectra and maps in Figure 2c in the main text were computed from the standard equation for the local rate of losses:

$$Q_{abs,local} = \langle \mathbf{j} \cdot \mathbf{E} \rangle_{time} = \mathrm{Im}(\varepsilon_{metal}) \frac{\omega}{8\pi} \mathbf{E}_{\omega} \cdot \mathbf{E}_{\omega}^{*}, \qquad \text{(S1)}$$

where $\mathbf{E}_{\omega}$ and $\varepsilon_{metal}$ are the complex field amplitude and the metal dielectric function, respectively.

<u>*Classical simulations of the plasmonic dynamics*</u>. Another important parameter of our trimer system is the splitting between the wavelengths of the symmetric (bright) and antisymmetric (dark) modes. The experimental and theoretical data in the figures in the main text reveal only the optically-allowed, symmetric modes, since the system is relatively small and the incident plane wave excites predominantly the symmetric modes. In theory, we can now excite the system with two dipoles that have opposite phases (Supplementary Figure S3). In this way, we create an antisymmetric driving field for the system and, therefore, the dark modes should become active in the extinction and absorption spectra. Then, we now clearly see the splitting between the symmetric (optically bright) mode and the antisymmetric (optically dark) mode in Figure S3. Using this splitting, we can estimate the characteristic transfer time in our trimer system. The splitting between the peaks for the two modes is equal to 0.22eV in energy units. We now assume that, using an ultra-short pulse, we excite only AuNP1. This thought experiment can be realized using an exciting plasmonic tip that is placed near AuNP1. Then, initially the excitation predominantly affects AuNP1. In other words, the ultra-short localized pulse sets up the initial conditions at $t = 0$ for the motion of the dipoles of the two AuNPs: $P_{NP1}(t = 0) = P_0$ and $P_{NP2}(t = 0) = 0$, where $P_{NP1}(t)$ and $P_{NP2}(t)$ are the polarizations of the AuNP1 and AuNP2, correspondingly. In this simplified approach, we can ignore the AgNP, which plays the role of a transmitter, and consider only the AuNPs regarding them as AuNP1 and AuNP2. The time dynamics of a harmonic system with two modes are the following, including the previously stated initial conditions:

$$P_{NP1}(t) = \frac{P_0}{2}\left(\cos[\omega_1 t] + \cos[\omega_2 t]\right) = P_0 \cos[\frac{\omega_1 + \omega_2}{2} t] \cdot \cos[\frac{\omega_1 - \omega_2}{2} t],$$

$$P_{NP2}(t) = \frac{P_0}{2}\left(\cos[\omega_1 t] - \cos[\omega_2 t]\right) = -P_0 \sin[\frac{\omega_1 + \omega_2}{2} t] \cdot \sin[\frac{\omega_1 - \omega_2}{2} t].$$

The polarizations have fast and slow components and the slow component oscillates with the circular frequency $(\omega_1 - \omega_2)/2 = \Delta_{splitt}/2$. Since the energy of a NP is proportional to the square of the polarization, the slow components give the evolution of the energies stored on the NPs in the system:

$$E_{NP1}(t) \propto \cos^2[\frac{\omega_1 - \omega_2}{2} t],$$

$$E_{NP2}(t) \propto \sin^2[\frac{\omega_1 - \omega_2}{2} t].$$

The characteristic transfer time can be estimated as the time when AuNP2 receives half of the initial energy from AuNP1. It happens when $\frac{\omega_1 - \omega_2}{2} t = \frac{\pi}{4}$. This equation sets up the transfer time $t_{transfer} = \frac{\pi}{2(\omega_1 - \omega_2)}$ which we give in the main text.

The above estimates for the transfer time were obtained analytically assuming a bi-harmonic relaxation.

In real systems the relaxation is, of course, more complex and a range of harmonics is involved. Now we compute such dynamics. The time-dynamics are now calculated using the Fourier transform of the Green's function for a single point dipole placed near AuNP1 (Figure S7); the dipole is defined as $d_0(t) \cdot \delta(\mathbf{r} - \mathbf{r}_d)$, where $\mathbf{r}_d$ is the position of the dipole. In this way, we can excite both Au-Au modes, the bright and dark ones. The time-dependent value of the moment of the exciting point dipole has the Gaussian shape:

$$d_0(t) = d_0 e^{-i\omega_0 t - \left(\frac{t}{\Delta t}\right)^2}.$$

Then, the dipole moments induced on the AuNPs should be calculated as Fourier integrals:

$$d_{NP}(t)=\int_{-\infty}^{\infty} e^{-i\omega\cdot t} d_0(\omega)\cdot G_{NP}(\omega)\cdot d\omega,$$

where $G_{NP}(\omega)$ is the Green function of a AuNP dipole in the frequency domain. This function is calculated as a response to a point dipole with a harmonically-varying moment $d_\omega(t)=e^{-i\omega t}$. In other words, a solution of the Maxwell's equation gives the dipole response $G_{NP}(\omega)\cdot e^{-i\omega t}$ if we excite the system with the dipole $d_\omega(t)=e^{-i\omega t}$. The function $d_0(\omega)$ is the Fourier transform from the exciting dipole:

$$d_0(\omega)=\frac{d_0}{2\sqrt{\pi}}\Delta t\cdot e^{-\frac{1}{4}\Delta t^2(\omega-\omega_0)^2}.$$

The dielectric function of metal is taken such that $\mathrm{Im}[\varepsilon_{Au}(\omega)]>0$ for $\omega>0$. In Figures S7, S8 and S9, we show various time traces for the Au-Ag-Au trimer and for the Au-Au dimer excited by a point dipole with an 8.3 fs-pulse ($\Delta t=5fs$).

Quantum theoretical calculations: As described in the main text, we can model our nanoparticles as three-dimensional quantum oscillators (Figure 5a). To develop such a model, we first write down the Hamiltonian in the coordinate representation and then we turn this Hamiltonian into the plasmonic second-quantization representation, which will correspond to a quantum picture of plasmons. The Hamiltonian of our system involving the dipole-dipole couplings between the NPs has the form:

$$\hat{H}=\sum_{\substack{\alpha=(i,\gamma)\\ i=1,2,3\\ \gamma=x,y,z}}\frac{\hat{p}_\alpha^2}{2m_{eff,i}}+\frac{k_i\hat{x}_\alpha^2}{2}+\frac{1}{\varepsilon_{eff}}\sum_{i>j}\frac{\mathbf{d}_i\cdot\mathbf{d}_j-3(\mathbf{d}_i\cdot\hat{\mathbf{z}})(\mathbf{d}_j\cdot\hat{\mathbf{z}})}{R_{ij}^3}, \quad \text{(S2)}$$

$$\mathbf{d}_i=e_{eff,i}\mathbf{r}_i,$$

where $\hat{p}_\alpha$ is the momentum operator for the oscillator $\alpha=(i,\gamma)$, where the index $i$ is the NP number ($i=1,2,3$) and $\gamma$ indicates the direction of the oscillation ($\gamma=x,y,z$). In this quantum calculation, we choose the z-axis along the trimer. Furthermore, $R_{ij}$ is the NP-NP distance, and $e_{eff,i}$ is the effective plasmonic charge of the i-NP. In our system, we have two Au NPs ($i=1,2$) and one Ag NP ($i=3$) and, correspondingly, $e_{eff,1}=e_{eff,2}=e_{Au}$ and $e_{eff,3}=e_{Ag}$. The coefficient $\varepsilon_{eff}$ is an effective dielectric constant of the matrix. In this model, three quantum plasmonic oscillators are embedded into an effective medium and coupled via the Coulomb dipole-dipole interaction. The first two terms are just the internal energy of the oscillators with the following frequencies:

$$\omega_{p,Au}=\omega_1=\omega_2=\sqrt{\frac{k_{Au}}{m_{eff,Au}}}\,,\ \omega_{p,Ag}=\omega_3=\sqrt{\frac{k_{Ag}}{m_{eff,Ag}}}\,. \quad \text{(S3)}$$

These oscillator frequencies should be matched with the plasmon energies of the isolated Au and Ag NPs: $\hbar\omega_{p,Au}=2.29eV$ and $\hbar\omega_{p,Ag}=2.8eV$. We now can quantize the oscillations by introducing the corresponding creation and annihilation operators of plasmons:[4]

$$\hat{x}_\alpha=\sqrt{\frac{\hbar}{2m_{eff,\alpha}}}\left(\hat{c}_\alpha+\hat{c}_\alpha^\dagger\right),\quad \hat{p}_\alpha=i\sqrt{\frac{m_{eff,\alpha}\hbar\omega_{p,\alpha}}{2}}\left(\hat{c}_\alpha-\hat{c}_\alpha^\dagger\right).$$

These operators leads to diagonalization of the first two terms in Eq. S2 and give the energy operator for the non-interacting system of plasmons,

$$\hat{H}_0=\sum_\alpha \hbar\omega_{p,\alpha}\left(\hat{c}_\alpha^\dagger\hat{c}_\alpha+\frac{1}{2}\right),$$

with the corresponding frequencies of the plasmonic excitations. Then, the total Hamiltonian of the coupled NPs reads:

$$\hat{H}=\hat{H}_0+\hat{H}_{\text{int}},$$

$$\hat{H}_{\text{int}}=\frac{1}{\varepsilon_{eff}}\sum_{i>j}\frac{\mathbf{d}_i\cdot\mathbf{d}_j-3(\mathbf{d}_i\cdot\hat{\mathbf{z}})(\mathbf{d}_j\cdot\hat{\mathbf{z}})}{R_{ij}^3}=\sum_{\substack{\alpha=(i,\gamma),\\ \beta=(j,\gamma')\\ i>j}} w_{\alpha\beta}\hat{c}_\alpha^\dagger\hat{c}_\beta+v_{\alpha\beta}\hat{c}_\alpha^\dagger\hat{c}_\beta^\dagger+h.c., \quad \text{(S4)}$$

$$w_{(i,x),(j,x)}=w_{(i,y),(j,y)}=\sqrt{\frac{\hbar}{2m_{eff,i}}}\sqrt{\frac{\hbar}{2m_{eff,j}}}\frac{\left(e_{eff,i}e_{eff,j}\right)}{\varepsilon_{eff}R_{ij}^3},$$

$$w_{(i,z),(j,z)}=-2\sqrt{\frac{\hbar}{2m_{eff,i}}}\sqrt{\frac{\hbar}{2m_{eff,j}}}\frac{\left(e_{eff,i}e_{eff,j}\right)}{\varepsilon_{eff}R_{ij}^3}.$$

The above operator includes the Coulomb-interaction terms between all three plasmonic quantum oscillations. The easiest way to proceed with the Hamiltonians (S2) and (S4) is to write down the equations of motion in the presence of a monochromatic external field $\mathbf{E} = \mathbf{E}_0 \cos[\omega t]$ for the average velocities of the oscillators:

$$\frac{d\hat{v}_\alpha}{dt} = \frac{i}{\hbar}[\hat{v}_\alpha, \hat{H}] + \hat{R}, \quad \hat{v}_\alpha = \frac{\hat{p}_\alpha}{m},$$

where $\hat{R}$ is the relaxation operator. Then, the above quantum equation will lead to the equations for the averaged velocities and coordinates:

$$\overline{v}_\alpha = \langle \Psi(t) | \frac{\hat{p}_\alpha}{m} | \Psi(t) \rangle, \ \overline{x}_\alpha = \langle \Psi(t) | \hat{x}_\alpha | \Psi(t) \rangle.$$

Then, by calculating at the eigen-frequencies of the system, we find the energies of the coupled plasmonic modes. Since the Hamiltonian is quadratic, the equations of motion for our system are simple and have the classical-mechanics form:

$$\begin{aligned} m_\alpha \frac{dv_\alpha}{dt} &= -k_\alpha x_\alpha + e_{eff,\alpha} E_{on\,\alpha} - m_\alpha \Gamma v_\alpha, \\ E_{on\,\alpha} &= E_0(t) + \sum_{\beta \neq \alpha} E_{\beta \to \alpha}(t), \end{aligned} \qquad \text{(S5)}$$

where $E_{on\,\alpha}$ is the total dynamic field acting on the $\alpha$-oscillator, $E_{\beta \to \alpha}(t)$ is the electric field created by the $\beta$-oscillator and acting on the $\alpha$-oscillator; the phenomenological rate $\Gamma = 1/\tau$ describes slow energy dissipation in the system. The equations are easy to solve using the symmetry of the system and the standard approach of complex variables. First, consider the incident field in the z-direction (the molecular axis), $\mathbf{E}_0 = \hat{z}\,\mathrm{Re}\, E_0 e^{-i\omega t}$. Then, the velocities have only z-components and should be found in the form of $\mathrm{Re}[v_{iz,\omega} e^{-i\omega t}]$. The system of equations for the complex amplitudes of the velocity follows from Eq. S5:

$$\left\{\begin{array}{l}(\omega^2-\omega_{p,Au}^2+i\omega\Gamma)v_{1z,\omega}+\Omega_{Au-Au}^2v_{2z,\omega}+\Omega_{Au-Ag,a}^2v_{3z,\omega}=\dfrac{\omega e_{eff,Au}}{m_{eff,Au}}iE_0\\ \Omega_{Au-Au}^2v_{1z,\omega}+(\omega^2-\omega_{p,Au}^2+i\omega\Gamma)v_{2z,\omega}+\Omega_{Au-Ag,a}^2v_{3z,\omega}=\dfrac{\omega e_{eff,Au}}{m_{eff,Au}}iE_0\\ \Omega_{Au-Ag,b}^2v_{1z,\omega}+\Omega_{Au-Ag,b}^2v_{2z,\omega}+(\omega^2-\omega_{p,Ag}^2+i\omega\Gamma)v_{3z,\omega}=\dfrac{\omega e_{eff,Ag}}{m_{eff,Ag}}iE_0\end{array}\right\},\quad \text{(S6)}$$

where the key interaction parameters are given by

$$\Omega_{Au-Au}^2=\frac{2e_{eff,Au}^2}{\varepsilon_{eff}m_{eff,Au}R_{12}^3},\quad \Omega_{Au-Ag,a}^2=\frac{2e_{eff,Au}e_{eff,Ag}}{\varepsilon_{eff}m_{eff,Au}R_{13}^3},\quad \Omega_{Au-Ag,b}^2=\frac{2e_{eff,Au}e_{eff,Ag}}{\varepsilon_{eff}m_{eff,Ag}R_{13}^3}.$$

Equation S6 determines the frequencies of plasmonic excitations in the coupled system of three oscillators. In the limit of small dissipation ($\Gamma\to 0$), the eigen-frequencies of the system are given by

$$\det\left\{\begin{array}{ccc}\omega^2-\omega_{p,Au}^2 & \Omega_{Au-Au}^2 & \Omega_{Au-Ag,a}^2\\ \Omega_{Au-Au}^2 & \omega^2-\omega_{p,Au}^2 & \Omega_{Au-Ag,a}^2\\ \Omega_{Au-Ag,b}^2 & \Omega_{Au-Ag,b}^2 & \omega^2-\omega_{p,Ag}^2\end{array}\right\}=0.$$

This equation has three positive solutions that correspond to the three longitudinal plasmons (L-modes) in our system:

$$\omega_{LD,Au-like}=\sqrt{\omega_{p,Au}^2+\Omega_{Au-Au}^2},$$

$$\omega_{LB,Ag-like}=\sqrt{\frac{\omega_{p,Au}^2+\omega_{p,Ag}^2-\Omega_{Au-Au}^2}{2}+\frac{1}{2}\sqrt{\left(\omega_{p,Ag}^2-\omega_{p,Au}^2\right)^2+8\Omega_{Au-Ag,a}^2\Omega_{Au-Ag,b}^2+\Omega_{Au-Au}^4+2\Omega_{Au-Au}^2\omega_{p,Ag}^2-2\Omega_{Au-Au}^2\omega_{p,Au}^2}},$$

$$\omega_{LB,Au-like}=\sqrt{\frac{\omega_{p,Au}^2+\omega_{p,Ag}^2-\Omega_{Au-Au}^2}{2}-\frac{1}{2}\sqrt{\left(\omega_{p,Ag}^2-\omega_{p,Au}^2\right)^2+8\Omega_{Au-Ag,a}^2\Omega_{Au-Ag,b}^2+\Omega_{Au-Au}^4+2\Omega_{Au-Au}^2\omega_{p,Ag}^2-2\Omega_{Au-Au}^2\omega_{p,Au}^2}}.$$

Two of the L-modes are bright (LB-modes) and one L-mode is dark (LD-mode). In this model, the Ag-like LB-mode experiences a blue shift, and the main Au-like LB-mode shows a red shift. The Au-like LD-mode, which is not active in our experiments, is blue-shifted. In a similar way, we can solve the geometry with the T-modes when $\mathbf{E}_0 \parallel \hat{\mathbf{y}}$ or $\mathbf{E}_0 \parallel \hat{\mathbf{x}}$. The results for these configurations can be obtained from the above equations using the following formal equations for the interaction parameters:

$$\Omega^2_{Au-Au} = -\frac{e^2_{eff,Au}}{\varepsilon_{eff} m_{eff,Au} R^3_{12}}, \quad \Omega^2_{Au-Ag,a} = -\frac{e_{eff,Au} e_{eff,Ag}}{\varepsilon_{eff} m_{eff,Au} R^3_{13}}, \quad \Omega^2_{Au-Ag,b} = -\frac{e_{eff,Au} e_{eff,Ag}}{\varepsilon_{eff} m_{eff,Ag} R^3_{13}}.$$

Figure 5 of the main text summarizes results obtained from the above quantum model. Here we list the model parameters used in our quantum calculations:

(1) Radii of the particles: $R_{Au} = 20nm$ and $R_{Ag} = 15nm$ ;

(2) Effective optical dielectric constant of the medium: $\varepsilon_{eff} = 2.15$;

(3) Effective charges and effective masses of the electrons in the NPs: $e_{eff,Au} = e \cdot N_{e,Au}$, $e_{eff,Ag} = e \cdot N_{e,Ag}$, $m_{eff,Au} = m_0 \cdot N_{e,Au}$ and $m_{eff,Ag} = m_0 \cdot N_{e,Ag}$.

(4) Effective numbers of free carriers in the NPs: $N_{e,Au} = 1.5 \cdot 10^6$ and $N_{e,Ag} = 0.6 \cdot 10^6$.

We note that the chosen numbers for $N_{e,Au}$ and $N_{e,Ag}$ are of the same order of magnitude as the calculated numbers of free electrons in the Au and Ag NPs. By taking such numbers, we obtain a reasonable agreement with the experiment (Figure 5). This is another supporting argument towards our simplified quantum model of plasmons.

Finally, we should note again that our quantum model is simplified and qualitative, but nevertheless describes the main effects. Three things omitted in our quantum model are: (1) The dielectric screening of the Ag-plasmons due to the core dielectric constant of the Au particles, (2) the interband transitions in gold, and (3) the multipole plasmon-plasmon interactions.

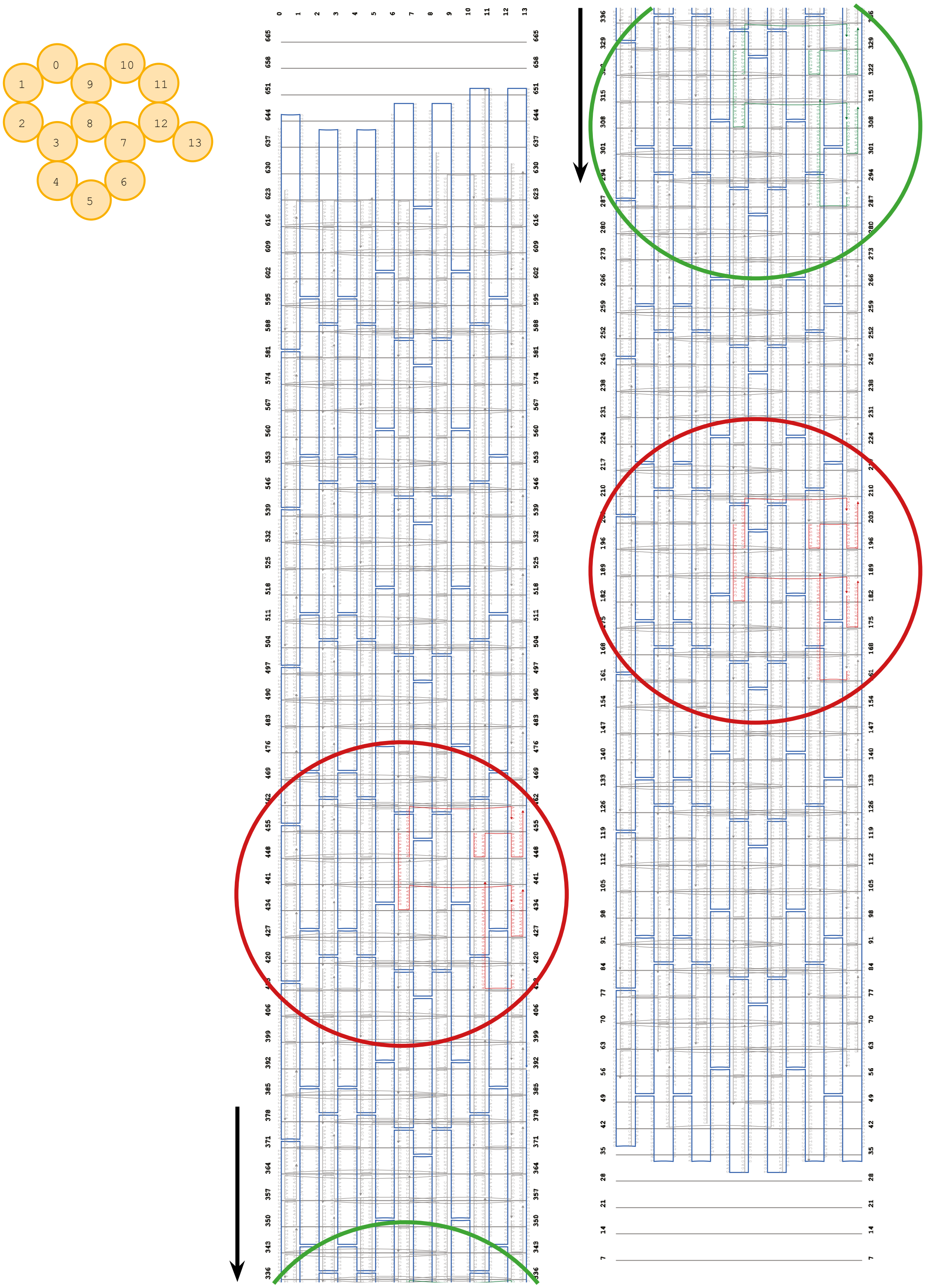

**Supplementary Figure S1. Design of 14hB DNA origami template structure.** The template structure consists of a 200 nm long 14-helix bundle. The scaffold path (blue),

the staple oligonucleotides (grey) and the elongated staple oligonucleotides at the nanoparticle attachment sites (red for AuNPs and green for AgNP) are displayed. The metal nanoparticle attachment is done *via* DNA hybridization of the functionalized AuNPs and AgNPs. The difference in the DNA sequence of the attachment site elongations as well as the complementary nanoparticle functionalizations guarantee site-specific binding.

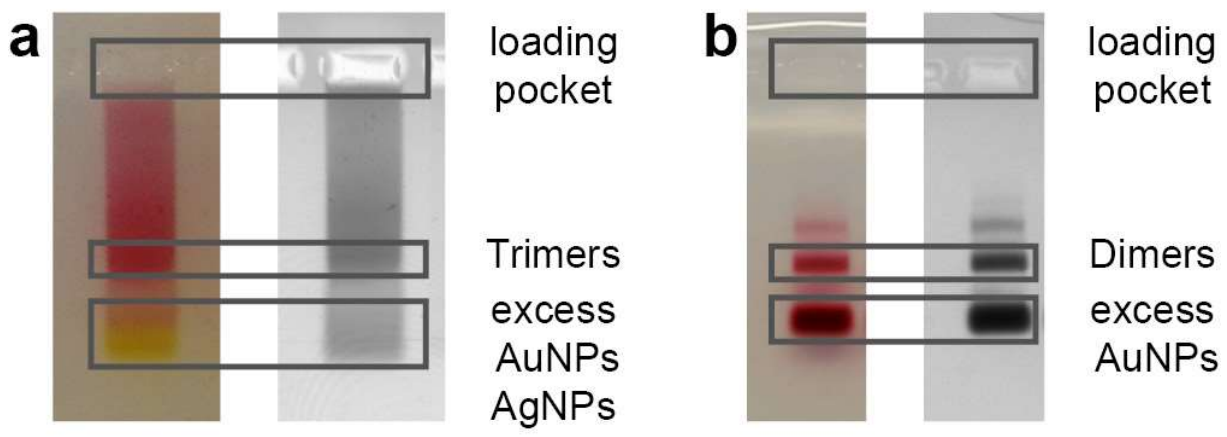

**Supplementary Figure S2. Gel purification of the nanoparticle-functionalized DNA origami structures. a,** Agarose gel purification of the trimer structure with two 40 nm AuNPs and one 30 nm AgNP attached to the DNA origami. **b,** gel purification of the dimer structure with two 40 nm AuNPs attached to the DNA origami.

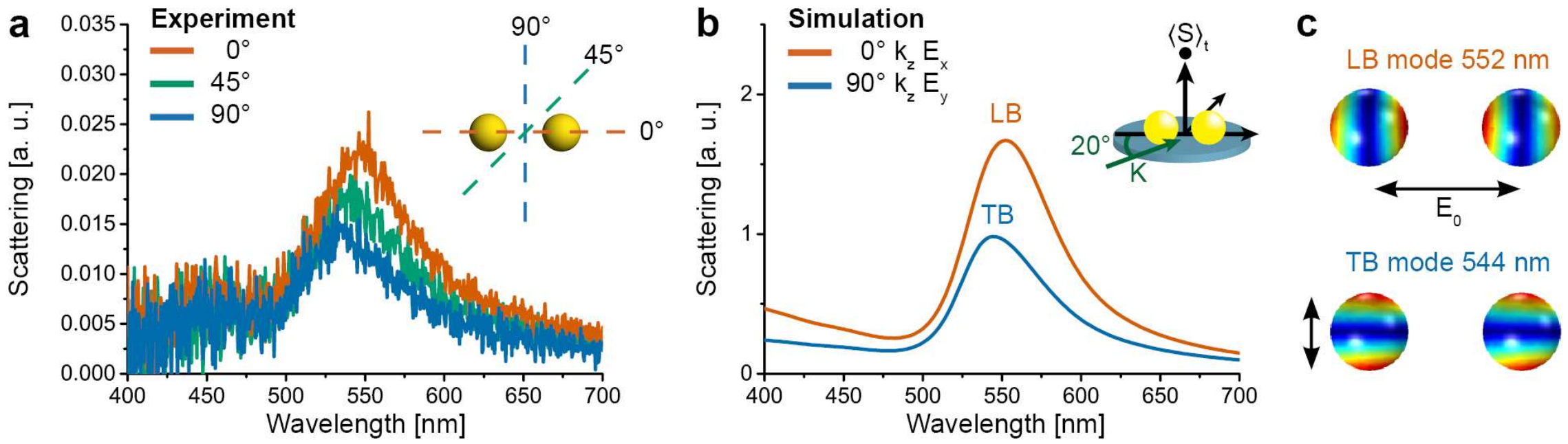

**Supplementary Figure S3. Polarization-dependent measurements and simulations of the homo-dimer structure. a,** Polarization resolved scattering spectra of a single AuNP dimer structure and **b,** corresponding simulations. Only a small peak shift between the longitudinal mode and the transversal mode is observed, reflecting the large distance

between the AuNPs so that they only couple slightly. Inset: The geometry used for calculations of scattering. **c,** The surface charge maps for the modes in the Au dimer.

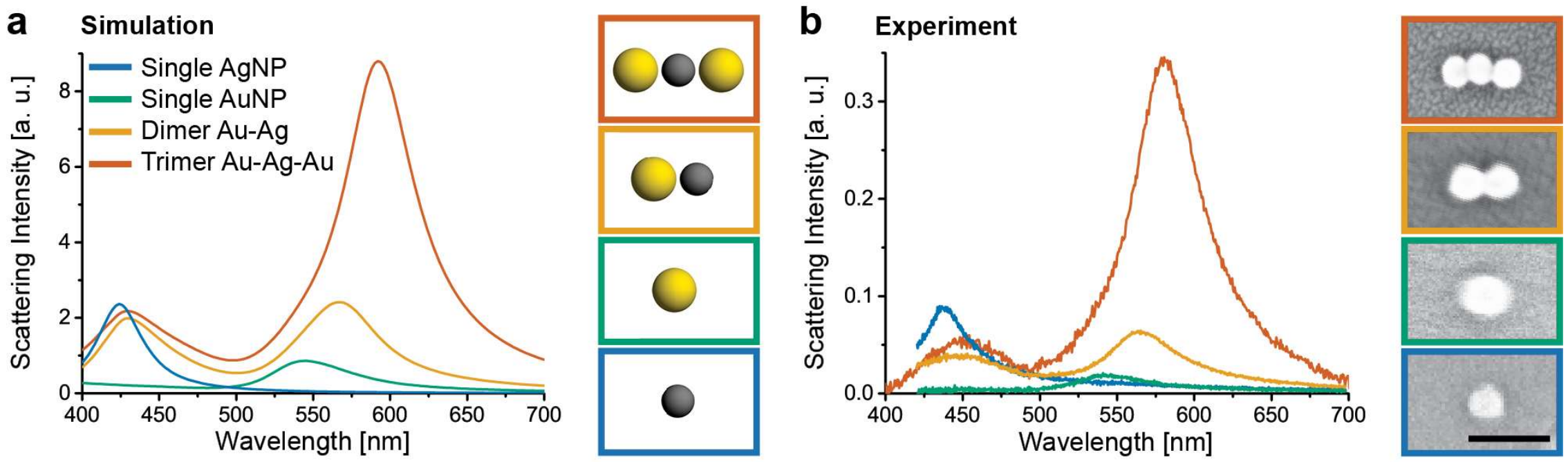

**Supplementary Figure S4. Optical properties of the Au-Ag dimer**. Simulations (**a**) and experimental spectra (**b**). In the Au-Ag dimer case,[4] the nearest-neighbor interaction in the Au-Ag pair creates a red shift and a two-fold enhancement of the scattering at the Au plasmon whereas, for the Au-Ag-Au trimer, a two times larger red shift and a much stronger enhancement (~ 10 times) of the Au-plasmon peak is observed. In the trimer case, there is a longer-range coupling and a strong Au-Au plasmon mediated by the AgNP is formed. These strong effects (the strong red shift and the 10-fold enhancement) indicate the formation of the coherent Au-Au plasmon excitation in the regime of loss-less plasmonic passage *via* the AgNP.

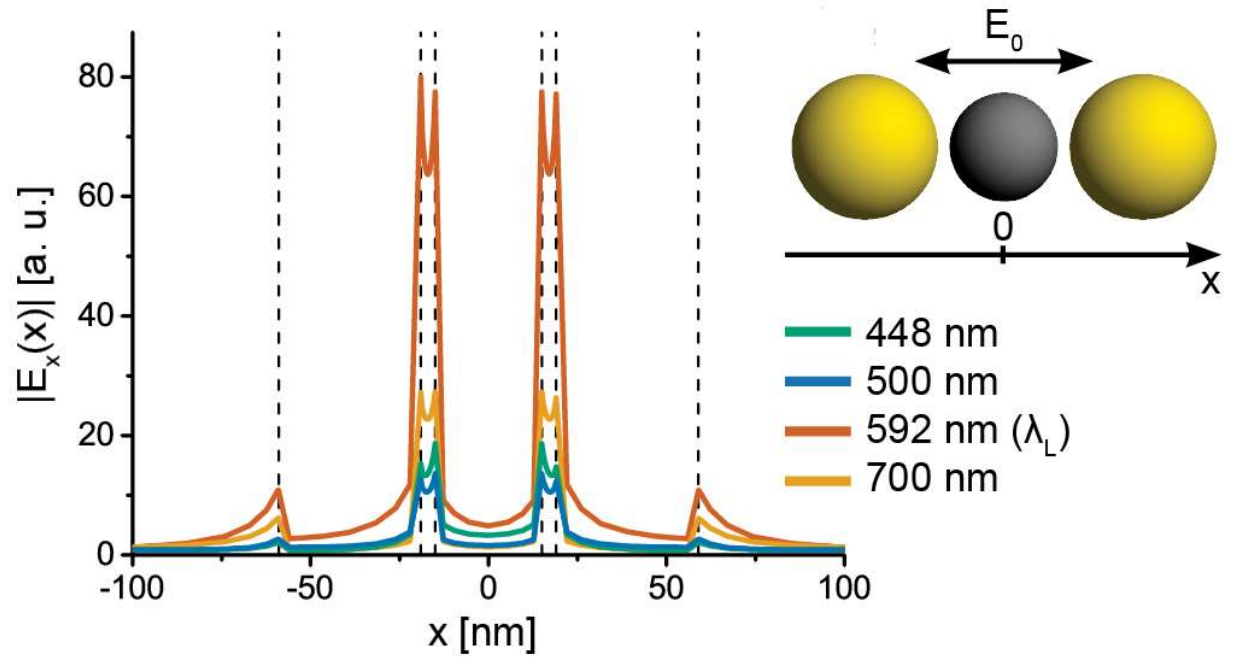

**Supplementary Figure S5. Hot Spots in the Au-Ag-Au trimer.** The hot spots in the trimer in terms of the local electric field plotted along the trimer axis are shown. The formation of very strong hot spots at the plasmon frequency of the bright Au-Au plasmon in the Au-Ag-Au trimer is observed. The electromagnetic hot spots are located in the gaps, where the electric fields become greatly enhanced. In the main text, we show energy dissipation and surface-charge maps that are fully consistent with this picture.

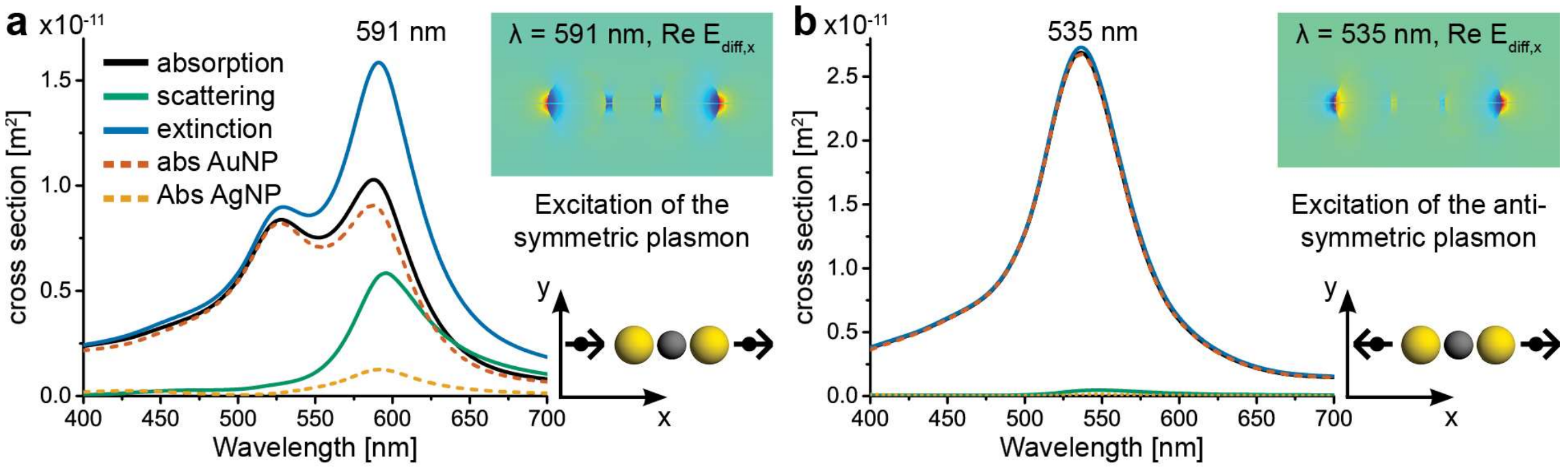

**Supplementary Figure S6. Model dipoles.** Calculations for the Au-Ag-Au trimer excited with two time-dependent dipoles placed on the left- and right-hand sides of our system. The panels show the field maps and the optical cross sections of our system under **a,** symmetric and **b,** anti-symmetric local excitations.

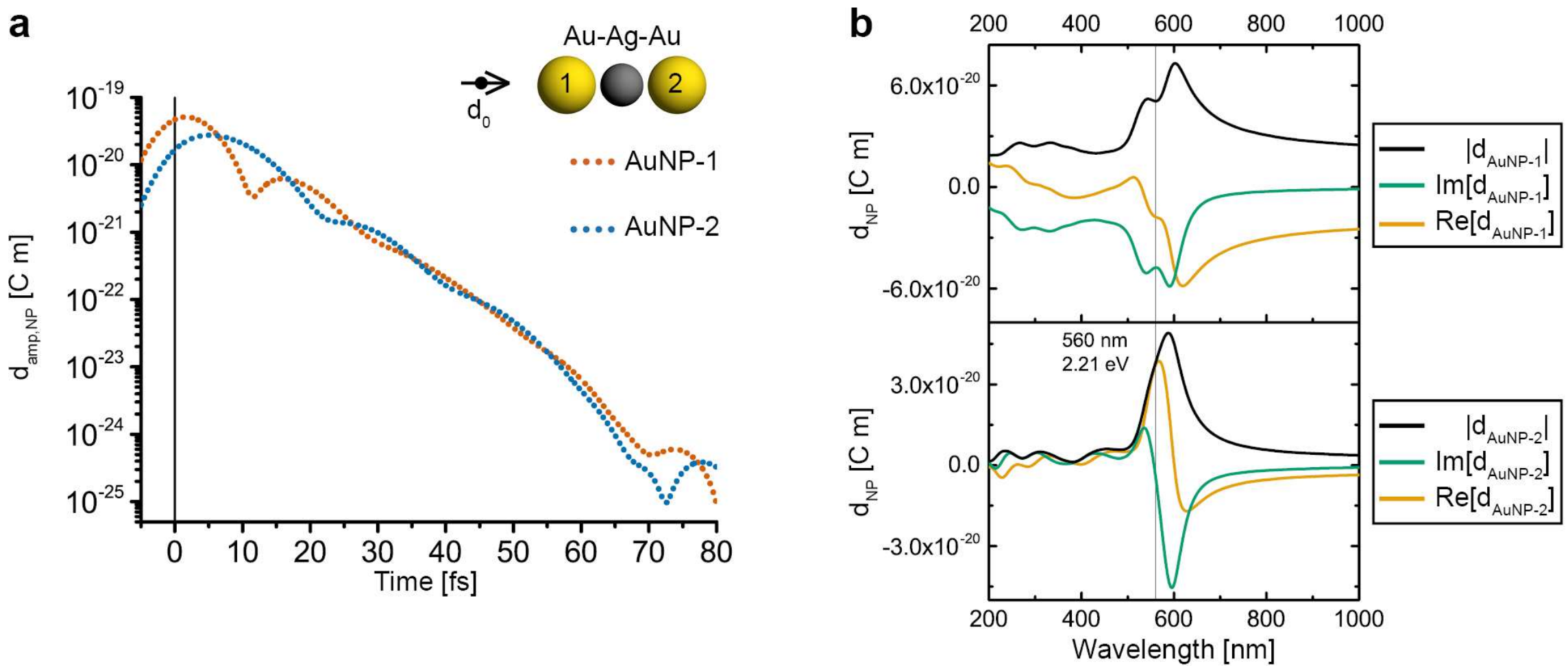

**Supplementary Figure S7. Analysis of the Au-Ag-Au trimer structure. a,** Time traces of the electric dipoles of the AuNPs in a logarithmic scale reveal coherent oscillations over long times. **b,** Spectral response functions of the excited dipoles of the AuNPs. These spectral functions show two peaks, which are the two major Au-Au plasmons (bright and dark modes). The beatings between these two modes in the panel (a) lead to the energy transfer effect and the splitting between these two modes gives an estimate for the transfer time in the hetero-trimer. The exciting dipole was taken as $d_0 = -1 \cdot 10^{-19} C \cdot m$ with a distance of 20 nm from the first AuNP. We are dealing here with the strong coupling regime since the splitting between the two major Au-Au modes (bright and dark) is clearly observed. Simultaneously, the time-dynamics shows the Rabi-like time beating between two AuNPs.

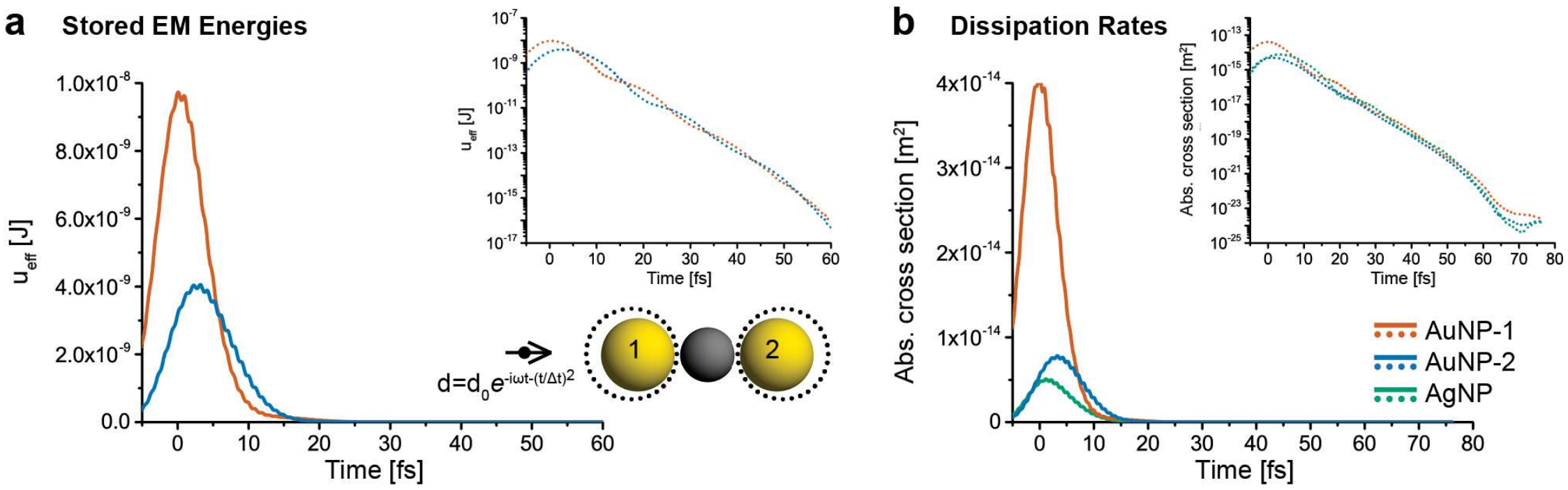

**Supplementary Figure S8. Time traces of the Au-Ag-Au trimer structure. a,** The electromagnetic energies stored near and inside the AuNPs. The spheres around the NPs show the integration regions. The electromagnetic energies of the NPs were calculated by taking integrals of the local electromagnetic energy density over the spheres including the AuNPs. **b,** Rates of dissipation in the three NPs displayed with linear and logarithmic scales. The dissipation rates inside the Au and Ag nanoparticles were computed by integration of Eq. S1 over the NP volumes.

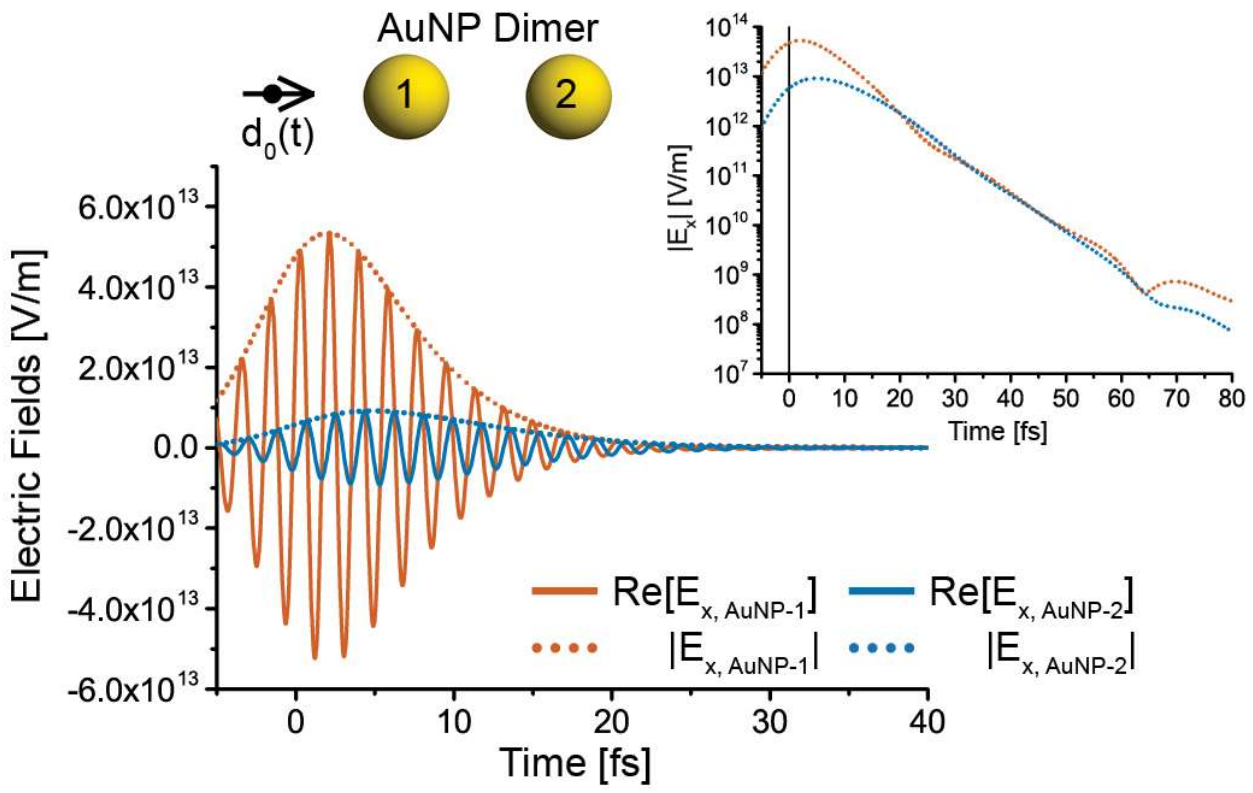

**Supplementary Figure S9. Time traces of the Au-Au dimer structure.** The time dynamics for the Au-Au dimer. This figure shows the electric fields in the centers of the AuNPs. For this system, we do not see essential beatings since the interaction in the Au-

Au dimer is relatively weak. Instead we mainly see the exponential relaxation due to Drude dissipation in the metals.